\documentstyle[aps,epsf]{revtex}

\newcommand{\hdmo}{Hei\-del\-berg-Moscow ex\-pe\-ri\-ment}

\def \be {\begin{equation}}
\def \ee {\end{equation}}
\def \ba {\begin{eqnarray}}
\def \ea {\end{eqnarray}}

\begin{document}

\title{First Results from the Heidelberg Dark Matter Search Experiment}

\author{L. Baudis \footnote{laura.baudis@mpi-hd.mpg.de}, A. Dietz,
  B. Majorovits, F. Schwamm, H. Strecker 
and H. V. Klapdor--Kleingrothaus \footnote{klapdor@daniel.mpi-hd.mpg.de}}

\address{Max-Planck-Institut f\"ur Kernphysik, Heidelberg, Germany}

\maketitle
\begin{abstract}
The Heidelberg Dark Matter Search Experiment (HDMS) is a new ionization
Germanium experiment in a special design. Two concentric Ge crystals
are housed by one cryostat system, the outer detector acting as an effective 
shield against multiple scattered photons for the inner crystal,
which is the actual dark matter target.
We present first results after successfully running the prototype
detector for a period of about 15 months in the Gran Sasso 
Underground Laboratory. We analyze the results in terms of 
limits on WIMP-nucleon cross sections and present the status of the
full scale experiment, which will be installed in Gran Sasso in the
course of this year.
\end{abstract}

\section{Introduction}

Weakly Interacting Massive Particles (WIMPs) are leading candidates
for the dominant form of matter in our Galaxy.
These hypothetical, relic particles from an early phase of the
Universe are predicted 
independently from cosmological considerations by supersymmetric 
particle physics theories as neutralinos - the lightest supersymmetric
particles.

Direct WIMP detection experiments exploit the elastic WIMP scattering 
off nuclei in a terrestrial detector \cite{goodwitt}.
However, detecting WIMPs is not a simple task. 
As their name suggests, their
interaction with matter is very feeble ($\sigma \leq \sigma_{weak}$) 
and predicted rates in supersymmetric models range from 10 to
10$^{-5}$ events per kilogram of detector material and day \cite{theo_rates}.
Moreover, for WIMP masses between a few GeV and 1\,TeV, the energy
deposited by the recoil nucleus is less then 100\,keV.
Thus, in order to be able to detect a WIMP, an experiment with 
a low energy threshold and an extremely low radioactive background 
is required.
Since the reward would be no less than discovering the dark matter in
the Universe, a huge effort is put into direct detection experiments.
More than a dozen of experiments are running at present and even more are
planned for the future (for recent reviews see \cite{morales,yorckdm,laura}).

The focus of this paper is to present first results of the Heidelberg
Dark Matter Search (HDMS) prototype \cite{hdms}, 
which took data over a period of about 15 months in the Gran Sasso 
Underground Laboratory (LNGS) in Italy.
After a description of the experimental setup, its
performance is discussed in some detail. The last 49 days of data taking
are analyzed in terms of WIMP-nucleon cross sections and a
comparison to other running dark matter experiments is made. 
The status and the prospects of the full scale
experiment are discussed and finally conclusions and an outlook are given.

\section{Description of the experiment}

\begin{figure}  
\centering 
\leavevmode\epsfxsize=200pt  
\epsfbox{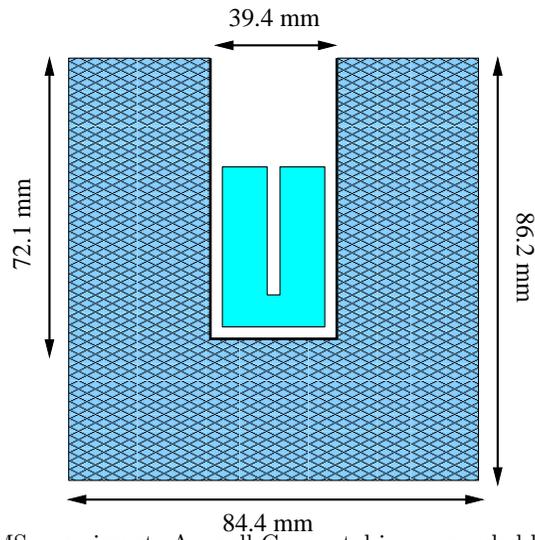}
\caption{\label{detindet} Schematic view of the HDMS experiment. A
  small Ge-crystal is surrounded by a well type Ge-crystal,
  the anticoincidence between them is used to suppress the background
  created by external photons.}
\end{figure}

HDMS operates two ionization HPGe detectors in a unique configuration \cite{hdms}.
A small, p-type Ge crystal is surrounded by a well-type Ge crystal,
both being mounted into a common cryostat system (see
Figure~\ref{detindet} for a schematic view). To shield leakage
currents on the surfaces, a 1\,mm thin insulator made from vespel is placed
between them. Two effects are expected to reduce the background of the
inner, WIMP  target detector with respect to our best measurements of the
Heidelberg-Moscow experiment \cite{ang2_prd}. First, the anticoincidence
between the two detectors acts as an effective suppression for multiple
scattered photons, second, the detection crystal is surrounded by Ge, 
which is one of the radio-purest known materials.
From previous measurements we know that the main radioactive
background of Ge detectors comes from materials situated in the immediate
vicinity of the crystals, i.e. from the copper parts of the cryostats
\cite{hdmo2000}.

\begin{figure}  
\centering 
\leavevmode\epsfxsize=200pt  
\epsfbox{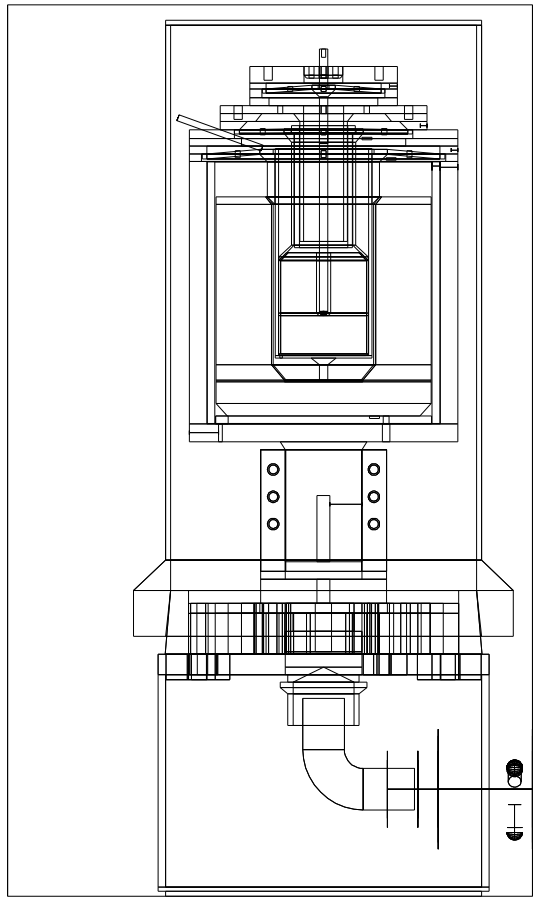}
\caption{\label{holder} Detailed view of the inner and
  outer Germanium crystals in their copper holder system.}
\end{figure}

In order to house both Ge crystals, a special design of the copper
crystal holder system was required.
The idea was to construct a pyramidal structure made of copper on top of
the crystals, allowing to establish the two HV and two
signal contacts, which are fixed by springs made from vespel (see Figure
\ref{holder}) \cite{greg}.  
In the prototype version three of the four contacts 
were unfortunately soldered by the manufacturer.
The cryostat system was fabricated in Heidelberg from low
radioactivity, electropolished copper.
The FETs are placed 20\,cm away from the crystals so that their effect on the
background is minimized by a small solid angle
 and by 10\,cm of copper shielding.
For the prototype experiment both inner and outer detector are made
of natural high-purity Germanium.

The data acquisition system of the HDMS experiment allows data
sampling either in a calibration mode (filling a histogram in a memory module, fast data
acquisition) or in an event-by-event mode where in addition 
the pulse-shape of an individual event can be recorded. 
The energy outputs of the preamplifiers are divided and amplified with two
different shaping time constants (3\,$\mu$s and 4\,$\mu$s) and different gains at the
spectroscopy amplifiers in order to
record low energy (from threshold to 400\,keV) and high energy
(from  70\,keV to 8\,MeV) spectra.
The spectra are measured with 13bit ADCs, which also release the
trigger for an event by a peak detect signal.
Further triggers are vetoed until the 
complete event information (arrival time with an accuracy of 100 $\mu$s,
energy and eventually the pulse shape) has been recorded.
The dead-time of 200 $\mu$s (without pulse shape) is negligible for a
typical event rate of 1.4$\times$10$^{-3}$ Hz in the outer detector.
The anticoincidence between the two detectors is performed off-line.

\section{Detector performance in Heidelberg}

First tests of detector performance were done in the Heidelberg low
level laboratory using standard
calibration sources like $^{133}$Ba, $^{60}$Co, $^{214}$Am and $^{152}$Eu-$^{228}$Th.
The sensitivity of the detectors in their geometrical arrangement
was tested by a scan measurement with a collimated $^{133}$Ba source.
The achieved energy resolutions and thresholds are shown in Table
\ref{detindet-data} along with other detector properties.

\begin{figure}  
\centering 
\leavevmode\epsfxsize=300pt  
\epsfbox{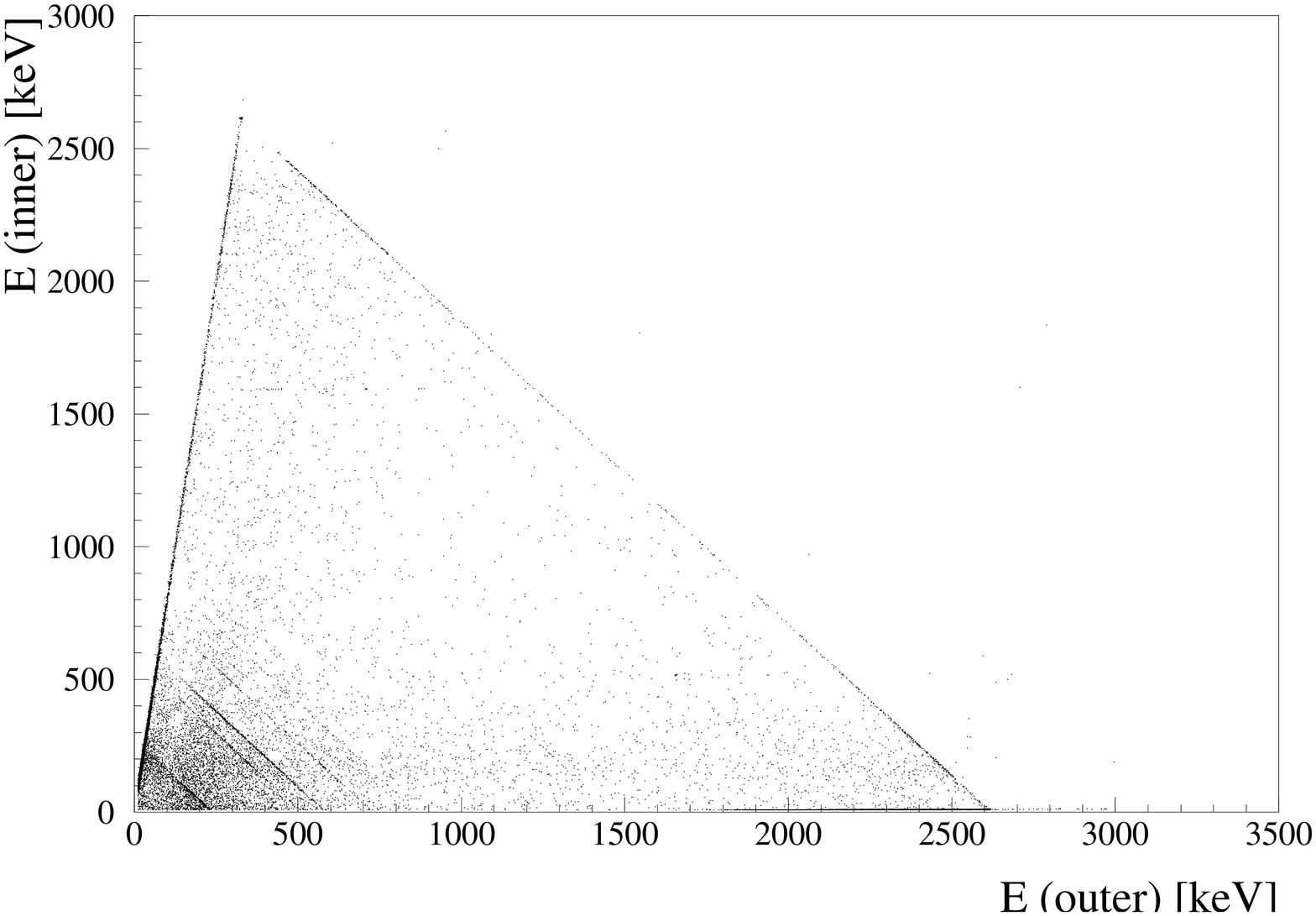}
\caption{\label{cross} Cross-talk between the inner and the outer Ge crystals,
measured using a $^{228}$Th source. 
The linear correlations are corrected off-line in order to obtain the true
anticoincidence spectra.}
\end{figure}

The concentric configuration of the two Ge-crystals 
gives rise to cross-talk events \cite{yorck-diss,npb70}.
Using $^{133}$Ba and $^{228}$Th calibration sources, we recorded data 
in an event-by-event mode, to obtain 
two-dimensional scatter plots of deposited energies in each detector. 
These plots clearly visualize a linear correlation between energies in
the inner and outer
crystal, although in different strengths (see Figure~\ref{cross}).

The linear correlation between the energy depositions in the two
crystals is corrected off-line.
If E$_i$  and E$_o$ are the true ionization energies deposited in the inner
and outer detector and E$^{\prime}_i$, E$^{\prime}_o$ the measured
energies, then one can write:
\ba
\nonumber
E_i = \frac{E^{\prime}_i - k_{io} \times E^{\prime}_o}{1 -
  k_{io}\times k_{oi}}\, ,\,\,\,\,\,\,\,\,\,\,\,\,\,
E_o = \frac{E^{\prime}_o - k_{oi} \times E^{\prime}_i}{1 - k_{io}\times k_{oi}},
\ea
where k$_{oi}$ and k$_{io}$ are the slopes of the correlation lines.
The width of the correlation lines is determined by the
energy resolution of the detectors. 
The slopes are determined to be:
\ba
\nonumber
k_{io} = 0.00375 \pm 0.43\times10^{-4}\\ 
\nonumber
k_{oi} = 0.12850 \pm 0.27\times10^{-3}
\ea 
The error in the energy determination arising from the cross-talk
correction is a function of the error of the measured energy
deposition and of 
the magnitude as well as the error of the slopes of the correlation lines.
It can be inferred directly by comparing the widths of energy peaks
in the two detectors after the cross-talk correction (for peaks 
with simultaneous energy deposition in the other detector, 
e.g. 511\,keV or the $^{60}$Co 
lines) with the widths of the same peaks for energy depositions in a 
single detector only.
While for the inner detector the effect is negligible (note that 
k$_{io}$ is much smaller than k$_{oi}$), for the outer   detector the
energy resolution worsens by 0.3\% to 0.4\% at 1332\,keV.
As we will see later, this has no influence on the
anticoincidence spectrum between the two detectors.  
The cross-talk correction requires the stability of the slopes of the linear
correlations. This stability has been monitored and confirmed. 
Another possibility to remove the cross-talk 
would be to introduce an extra grounded shield between the
detectors. We refrained from applying this hardware solution because
of strongly reduced Anti-Compton capability and the additional
contamination risk.

\section{Detector performance at LNGS}

\begin{figure}  
\centering 
\leavevmode\epsfxsize=300pt  
\epsfbox{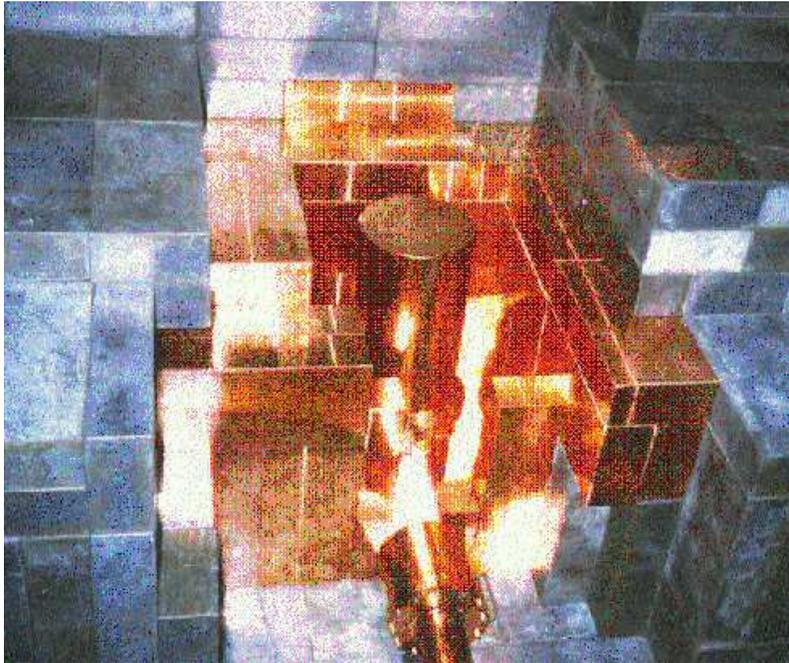}
\caption{\label{hdms-gs}The HDMS detector in its open shield 
during the installation in the Gran Sasso Underground Laboratory.
The inner shield is made of 10\,cm of electrolytic copper, the outer
one of 20\,cm of Boliden lead.}
\end{figure}

The HDMS prototype was installed at LNGS in March 1998 and successfully
took data over a period of about 15 month, until July 1999 \cite{diss}.
Figure \ref{hdms-gs} shows the detector in its open shield. The
inner shield is made of 10\,cm of electrolytic copper, the outer
one of 20\,cm of Boliden lead, both lead and copper having been stored 
for several years below ground at Gran Sasso.  The whole setup is enclosed in an air
tight steel box and flushed with gaseous nitrogen in order to suppress
environmental radon diffusion. Finally a 15 cm thick borated
polyethylene shield surrounds the steel box to minimize the
influence of neutrons.

The individual runs were about 0.9 d long. The experiment
was stopped daily and parameters such as leakage currents of the detectors,
nitrogen flow, overall trigger rate and count rate of the individual
 detectors were checked. 

\begin{figure}  
\centering 
\leavevmode\epsfxsize=300pt  
\epsfbox{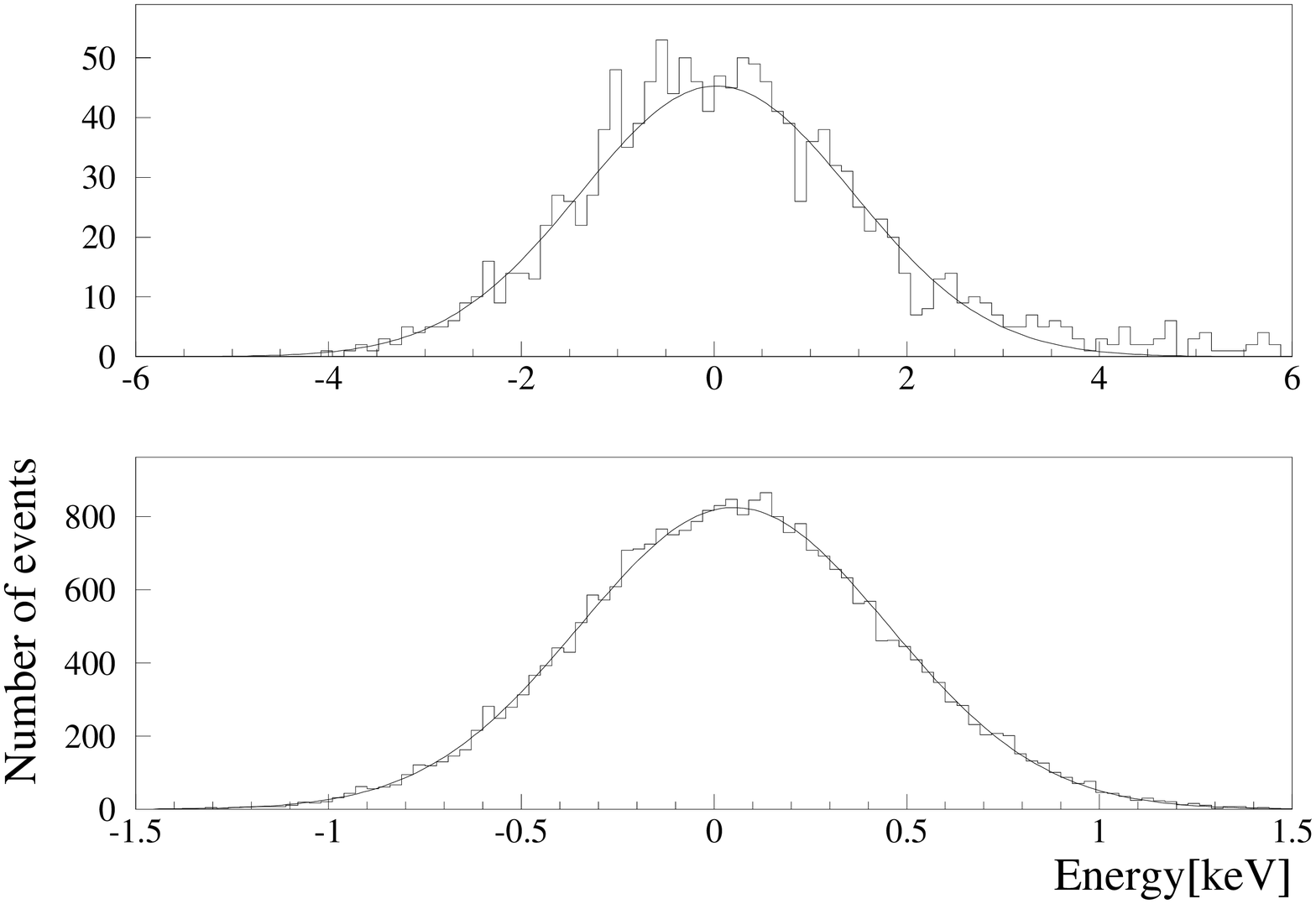}
\caption{\label{nullpkt}Zero energy resolution of the outer (upper)
  and inner (lower) detector, determined from the cross-talk
  correction. The widths of the Gauss-curves are
  1.42~keV and 0.4~keV, respectively.}
\end{figure}

Energy calibrations were done weekly with standard $^{133}$Ba
and $^{152}$Eu-$^{228}$Th sources, introduced through a Teflon tube. 
The energy resolution of both detectors (1.2\,keV at 300\,keV inner
detector and 3.2\,keV at 300\,keV outer detector) were stable as a
function of time. 
The zero energy resolutions, which  are determined by extrapolating 
the full widths at half maximum (FWHM) to 0\,keV, are
(1.06$\pm$0.3)\,keV and (3.04$\pm$0.3)\,keV for the inner and outer
detector, respectively.
Another way to specify the zero energy resolution is given by the
cross talk correction. The cross talk generates `fake' events beyond the
energy thresholds of both detectors, which are then shifted to 0\,keV
by the correction. This allows a direct measurement of the energy 
resolution at 0\,keV (see Fig. \ref{nullpkt}) and  a cross check of the
quality of the cross-talk correction at the same time. 
The values of 0.94\,keV and 3.34\,keV 
are fully consistent with the ones obtained by the FWHM-extrapolation.

To determine the energy threshold, a reliable energy calibration at low
energies is required. The lowest-energetic lines observed in the 
detectors in the calibration mode were the $^{133}$Ba line at 81\,keV
in the inner detector and the 121.8\,keV line of  $^{152}$Eu in the
outer one (all lower energetic lines are absorbed in the copper
of the crystal holder system).  However, intrinsic radioactivity of
the crystals and/or of the surrounding copper such as the 10.37\,keV 
line of  $^{68}$Ge and the 46.5\,keV line of $^{210}$Pb 
allowed to check the calibration down to low energies.
The energy thresholds are 2.0\,keV and 7.5\,keV for the inner and outer
detector, respectively. 

\begin{figure}  
\centering 
\leavevmode\epsfxsize=400pt  
\epsfbox{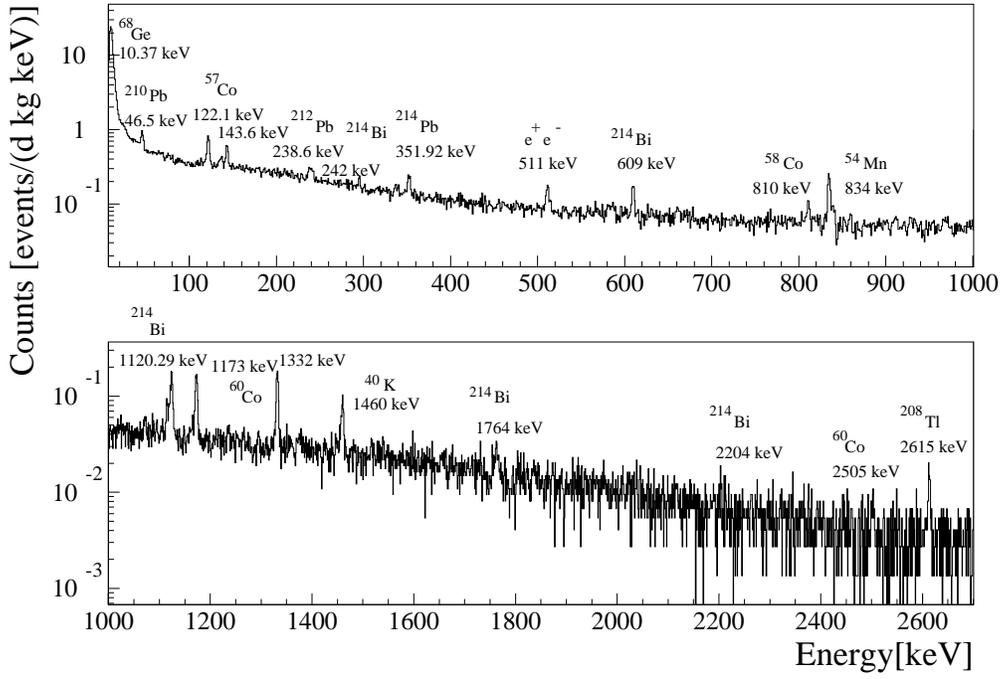}
\caption{\label{sum-outer} Sum spectrum of the outer detector after 
363 raw live-days of data taking. The most prominent lines are labeled.}
\end{figure}

\begin{figure}  
\centering 
\leavevmode\epsfxsize=400pt  
\epsfbox{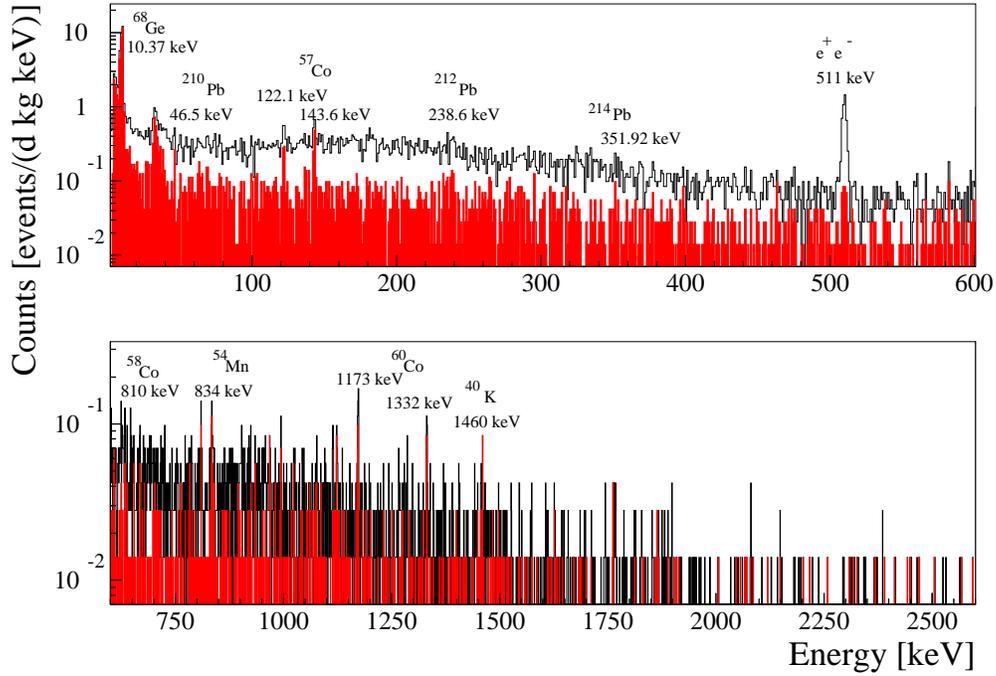}
\caption{\label{sum-inner} Sum spectrum of the inner detector after 
  363 raw live-days of data taking. The most prominent lines are
  labeled. The filled histogram is the resulting spectrum after
  computing the anticoincidence with the outer detector.}
\end{figure}

After correction for the cross talk and re-calibration to standard
calibration values (according to the weekly determined calibration
parameters), the spectra of the daily runs were summed up. Figures
\ref{sum-outer}  and  \ref{sum-inner} show the sum spectra for the
outer and inner detector, respectively. The most important identified 
lines are labeled.

\begin{figure}  
\centering 
\leavevmode\epsfxsize=300pt  
\epsfbox{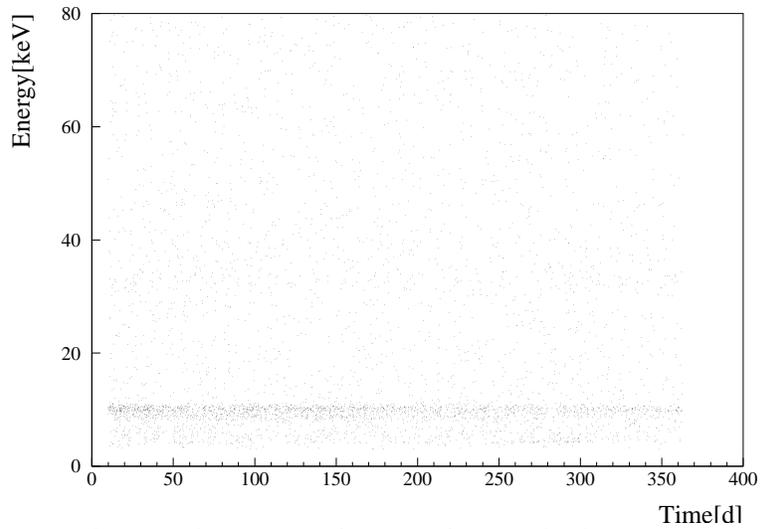}
\caption{\label{time-inner} Energy depositions in the inner detector
  as a function of time. The decreasing activity of $^{68}$Ge is
  nicely visualized. No microphony (bursts)  beyond
  2\,keV can be seen.}
\end{figure}

In the outer detector the lines of the cosmogenic isotopes 
$^{68}$Ge, $^{57}$Co, $^{58}$Co, $^{54}$Mn, $^{60}$Co, $^{65}$Zn,  
of the natural decay chains $^{238}$U and $^{232}$Th, 
of the primordial $^{40}$K, of the anthropogenic 
radionuclide $^{137}$Cs and the annihilation line at 511\,keV can be
identified. The statistics in the inner detector is lower, 
however the following lines can be clearly seen: $^{68}$Ga  K$_{\beta}$ 
at 10.37\,keV, $^{210}$Pb  at 46.5\,keV, external $^{57}$Co 
at 122.1\,keV and  internal at 143.6\,keV (+ X-ray), 511\,keV annihilation,
$^{54}$Mn at 834\,keV, $^{60}$Co at 1173\,keV and
1332\,keV and  $^{40}$K  at 1460\,keV.
The region below 10\,keV is dominated by the X-rays of  
$^{49}$Ti (5\,keV) and $^{55}$Mn (6.5\,keV). 
In addition to the outer detector, a
structure centered at 32\,keV with a FWHM of 2\,keV is identified. 
Its
origin is not yet fully understood, but could be a $^{210}$Pb
contamination at the inner contact with incomplete charge collection.
Figure \ref{time-inner} shows the energy depositions in the inner
detector as a function of measuring time. No microphonic events
(bursts) can be seen beyond 2.0\,keV. The decreasing activity of $^{68}$Ge is 
nicely visualized. 

After 363 days of pure measuring time the statistics in the inner
detector was high enough in order to estimate the background
reduction through the anticoincidence with the outer detector. In
order to obtain true anticoincidence spectra, the
energy thresholds for computing the anticoincidence must lie at least 
3$\sigma$ away from zero, where FWHM$_0$=2.35$\sigma$ is the zero energy 
resolution. Table \ref{nullpkt_tab} gives the energy thresholds
and the zero energy resolutions for both detectors, after the cross
talk correction. 
The energy thresholds lie in both cases about 5$\sigma$ away from
zero, so they count as cuts for computing the anticoincidence between
the crystals.  
Thus, the fact that the zero energy resolution of the outer detector
is worsened by the cross-talk correction does not affect the 
anticoincidence spectra between the two detectors.

\begin{figure}[t!]  
\centering 
\leavevmode\epsfxsize=260pt  
\epsfbox{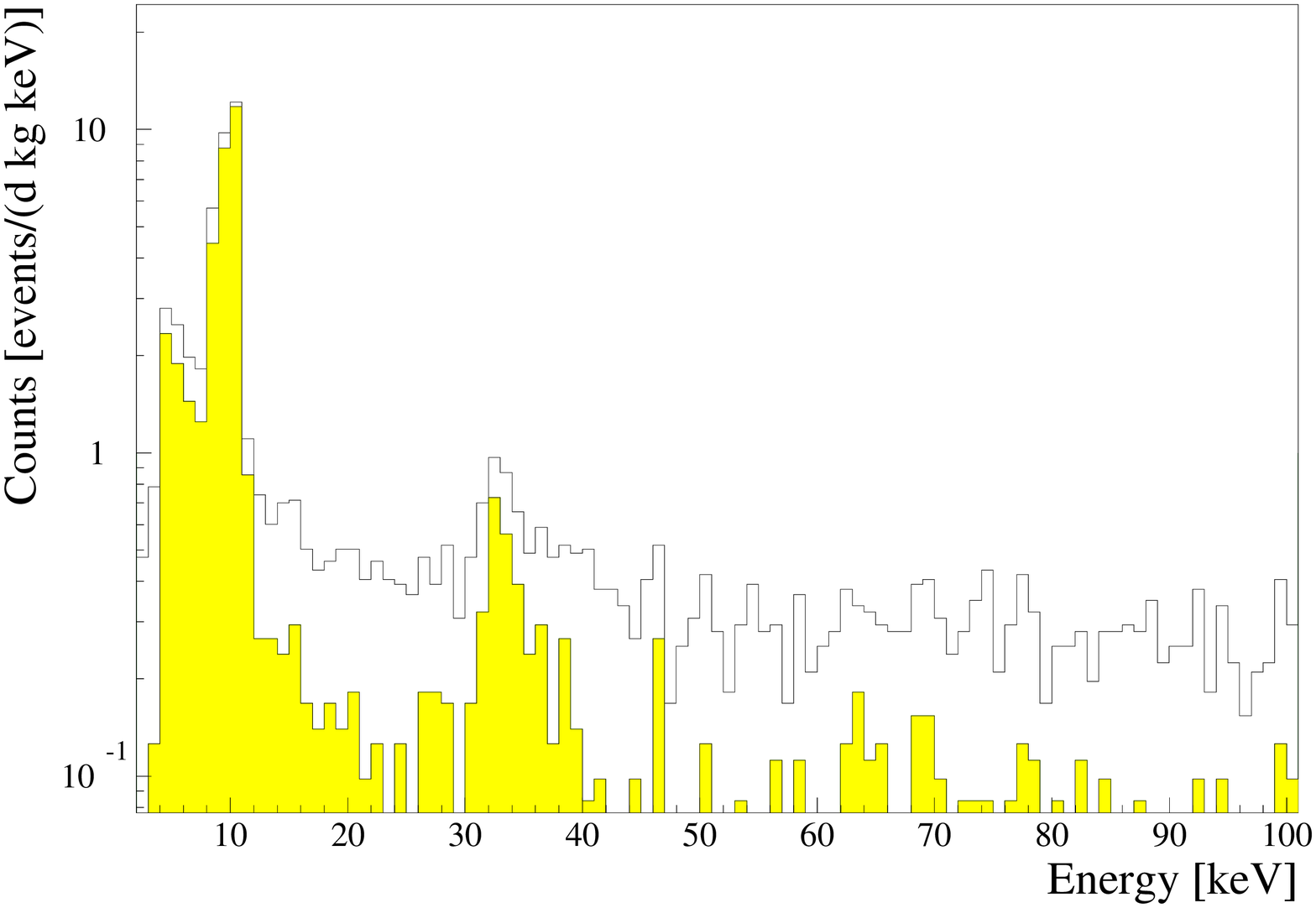}
\caption{\label{low-inner} Low energy spectrum of the inner, natural Ge detector
  before and after (filled histogram) the anticoincidence is applied with the
  outer Ge detector. The internal, low energy X-rays, as well as the
  structure centred at 32\,keV are not removed by the anticoincidence
  with the outer detector. A $^3$H $\beta^-$-spectrum with endpoint at 
  18\,keV is most likely present.}
\end{figure}

Figure \ref{low-inner} shows the low-energy spectrum of the inner detector
before and after the anticoincidence. 
The anticoincidence has no influence on the cosmogenic X-rays below 11\,keV, as well as 
on the structure at 32\,keV. The $\beta^-$spectrum from the cosmogenically 
produced $^3$H with endpoint at 18\,keV is most likely present as well.
If the Compton suppression is applied in the energy region between
40\,keV and 100\,keV, the background reduction factor is 4.3. 
The counting rate after the anticoincidence in this energy region is 
0.07\,events/kg\,d\,keV, thus very close to the value obtained in the 
Heidelberg-Moscow experiment with the enriched detector ANG2 \cite{ang2_prd}.    
In the energy region between 11\,keV and 40\,keV the background index is 
a factor of 3 higher (0.2\,events/kg\,d\,keV).

\section{Dark Matter Limits}

The evaluation for dark matter limits on the WIMP--nucleon cross section 
$\sigma_{\rm scalar}^{\rm W-N}$ follows the method described in \cite{ang2_prd}.
Because the cosmogenic radionuclides produced in the Ge crystals and
in the surrounding copper have typical half-lives of 300 days, 
we consider only the last 49\,d of measurement of the HDMS prototype. 
The total exposure corresponds to 9.9\,kg\,d. The number of counts per
1\,keV energy bin are listed in Table \ref{inner-counts}. 
The background index in the energy region between 2--30\,keV is 
0.5\,events/kg\,d\,keV. No background subtractions were applied.
The parameters used in the calculation of expected WIMP spectra are 
given in Table ~\ref{tab:parameter}.

The resulting upper limit exclusion plot in the 
$\sigma_{\rm{scalar}}^{\rm{W-N}}$ versus M$_{\rm{WIMP}}$
plane is shown in Fig.~\ref{dm_limits}.
At this stage, the limit is not yet competitive with our limit from
the Heidelberg-Moscow experiment for large WIMP masses.
However, the cross section limits for WIMP masses below 40\,GeV are 
considerably improved. This is mainly due to the lower energy
threshold of 2\,keV, compared to 9\,keV in the Heidelberg-Moscow
experiment. 
Also shown in the figure are limits from the Heidelberg-Moscow
experiment \cite{ang2_prd}, the Neuch$\hat{\rm a}$tel-experiment \cite{neuchat},
the DAMA experiment \cite{dama} and the most recent limits from the 
CDMS experiment \cite{rick2000}.
The  filled contour represents  the 2$\sigma$ evidence region of the DAMA
  experiment \cite{dama3}.    
The dashed line is the expectation for HDMS having a background index
equal to the one of the Heidelberg-Moscow experiment \cite{ang2_prd}
extended down to an energy threshold of 2\,keV.
The experimental limits are compared to
expectations (scatter plot) for WIMP-neutralinos calculated in the
MSSM parameter space at the weak scale (without any GUT constraints)
under the assumption that all superpartner masses are lower than
300 GeV - 400 GeV \cite{vadim99}.

\section{The full scale HDMS experiment}

For the second phase of HDMS, important  changes were already made.
First, 
the inner crystal made of natural Germanium was replaced by 
an enriched $^{73}$Ge crystal. In this way, we will be sensitive also
to spin-dependent WIMP-nucleon interactions, since $^{73}$Ge is the
only Germanium isotope with spin.
In addition,  the $^{70}$Ge isotope, which is the source of
$^{68}$Ge by the $^{70}$Ge(p,t)$^{68}$Ge reaction, is now strongly
depleted. Although the exact depletion is not measured yet,
we expect at least a factor of 50 (we measured a factor of 60
depletion in  $^{70}$Ge for the enriched $^{76}$Ge detectors in
the {\hdmo} \cite{jochen-diplom}). 
 
Second,
the copper crystal holder system  was replaced by a holder made of extremely
radio-pure copper; soldering of the signal and the high-voltage
contacts was not applied.
We believe that both facts will have a large impact on the background of the HDMS
detectors, since Monte Carlo simulations based on {\sc Geant3.21}
showed \cite{frank-diplom}, that besides the cosmogenic activation 
of the crystals, the U/Th contamination of the copper and solder 
 were the main background sources of the prototype.

The $^{73}$Ge-HDMS detector has been assembled  and 
transported by ship to Heidelberg, where its performance was tested 
in the low-level laboratory. Both inner and outer
detector are working well, the (not optimized) energy resolutions
being around 1.9\,keV and 4.0\,keV at 1332\,keV.
We will install  the full scale experiment at LNGS in August 2000.

\section{Summary and Outlook}

The prototype detector of the HDMS experiment successfully took data 
at LNGS over a period of about 15 months. 
Most of the dominant background sources  were identified.
However, the origin of a structure centred at 32\,keV in the inner
detector is still unclear. Due to its low energy and the fact that it is not removed
by the anticoincidence with the outer detector, it is most probably 
located in the inner crystal itself or at the signal contact. 
The future measurement with the $^{73}$Ge-HDMS detector will hopefully 
help to clarify its nature. 

The background reduction factor through anticoincidence
 in the inner detector in the energy
region 40--100\,keV is  4.3. It is 
less then previously expected \cite{hdms}, due to the smaller
diameter of the outer, veto Ge-detector than originally planned.
Nevertheless, the background in the energy region between 40\,keV and
100\,keV of the inner detector 
is with 0.07\,events/kg\,d\,keV already now at the level of the Heidelberg-Moscow
experiment. In the region between 2--30\,keV it is with
0.5\,events/kg\,d\,keV a factor of 7 higher, mainly due to the
cosmogenic activation of the natural Ge crystals. However, 
due to the smaller energy threshold of 2\,keV, the limits on 
WIMP-nucleon cross sections were improved for WIMP masses below
40\,GeV with respect to the Heidelberg-Moscow experiment.
They are currently the most stringent spin-independent limits at low 
WIMP masses for using raw data without any background subtraction.

\begin{figure}
\centering 
\leavevmode\epsfxsize=300pt  
\epsfbox{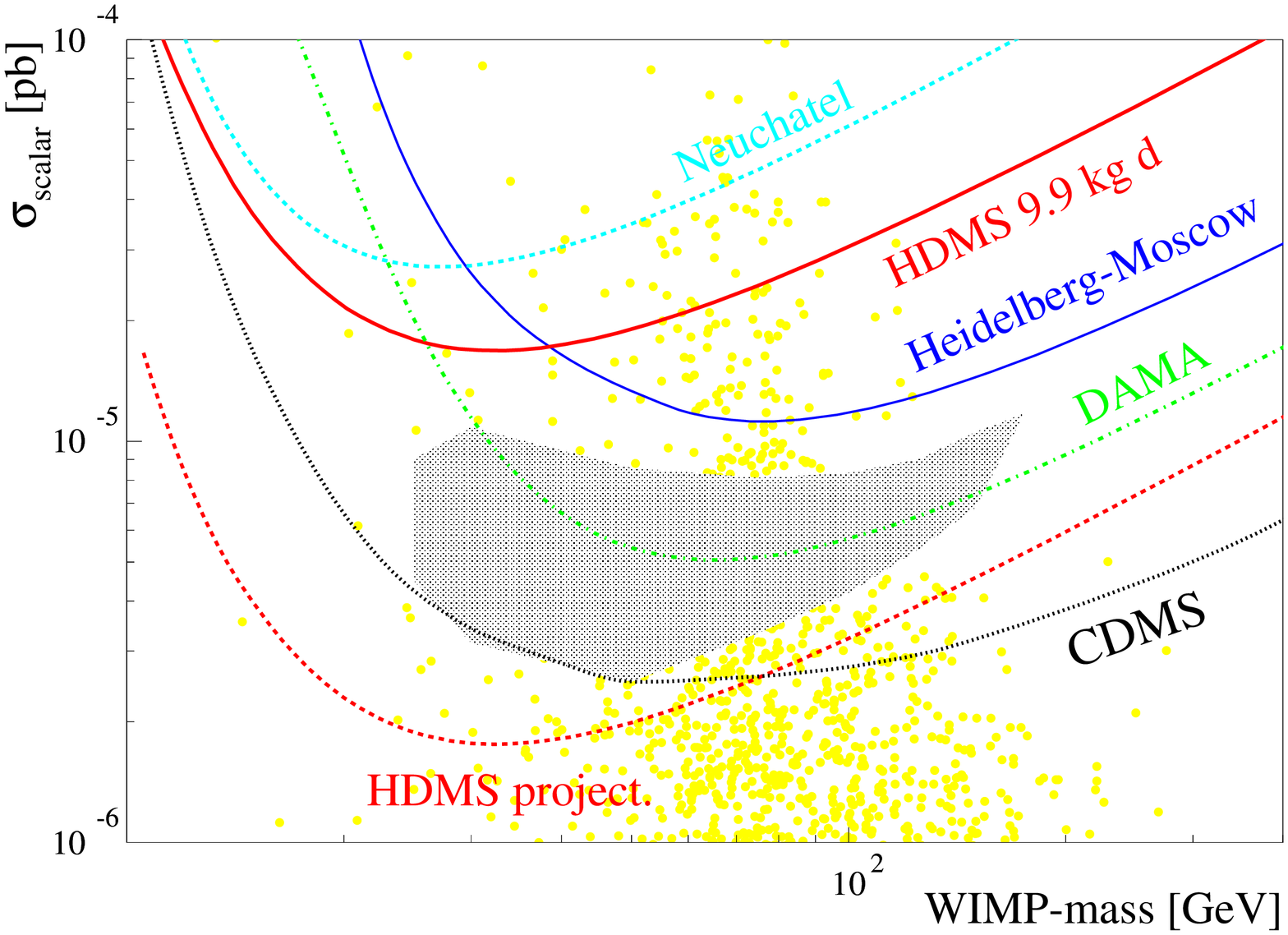}
\caption{
WIMP-nucleon cross section limits as a function of the WIMP
  mass for spin-independent interactions.
Solid thick line: 90\% C.L. limit of the HDMS prototype with 
9.9\,kg\,d exposure.
Solid thin line: limit of the Heidelberg-Moscow experiment
\protect{\cite{ang2_prd}}, light dashed line: Ge limit from Neuch$\hat{\rm a}$tel
\protect{\cite{neuchat}}, dashed-dotted line: DAMA limit \protect{\cite{dama}},
dotted line: CDMS 2000 limit \protect{\cite{rick2000}}. 
The dark dashed curve is the expectation for HDMS with a background 
index of 0.07 events/kg\,d\,keV in the 2-30\,keV energy region.
The  filled contour represents  the 2$\sigma$ evidence region of the DAMA
  experiment \protect{\cite{dama3}}.    
The experimental limits are compared to
expectations (scatter plot) for WIMP-neutralinos calculated in the
MSSM parameter space at the weak scale (without any GUT constraints)
under the assumption that all superpartner masses are lower than
300 GeV - 400 GeV \protect{\cite{vadim99}}.}
\label{dm_limits}
\end{figure}

The expectation for the $^{73}$Ge-HDMS detector is shown in
Fig.~\ref{dm_limits}. It is based on a  background index of 
0.07\,events/kg\,d\,keV in the region between 2--30\,keV.
The aim of HDMS is to test the `evidence region' singled out by the DAMA
experiment \cite{dama3} in the MSSM parameter space 
after about two years of measurement. It would be an independent test
by using only raw data and a completely different detection technique.

\acknowledgements
We would like to thank Dr. Gerd Heusser and Dr. Yorck Ramachers for 
valuable discussions.

\begin{table}
\caption{Detector properties for the small inner Ge-detector and the active veto-shield, the outer well-type Ge-detector.}
\label{detindet-data}
\begin{tabular}{lcc}
Property & Inner Detector & Outer Detector\\
\hline
Crystal Type & p--type & n--type \\
Mass [g] & 202 & 2111\\
Active Volume [cc] & 37 & 383 \\
Crystal diameter [mm] & 35.2 & 84.4 \\
Crystal length [mm] & 40.3 & 86.2 \\
Operation Bias & +2500 & -1500 \\
FWHM (1332\,keV) [keV] (ORTEC) & 1.92 & 4.41 \\
FWHM (1332\,keV) [keV] (Heidelberg) & 1.87 & 4.45 \\
Threshold Heidelberg [keV] & 2.5 & 7.5 \\
Optimum pulse shaping [$\mu$s] & 4 & 2--3 \\
\end{tabular}
\end{table}

\newpage

\begin{table}
\caption{Electronic energy thresholds and zero energy resolutions for the inner and outer Ge-detector.}
\begin{tabular}{lcc}
&  threshold [keV] & $\sigma_0$ = FWHM$_0$/2.35\\
\hline
inner & 2.0 & 0.40 \\
outer& 7.5 & 1.42 \\
\end{tabular}
\label{nullpkt_tab}
\end{table}

\begin{table}
\caption{Number of counts per 1 keV energy bin after an exposure of 9.9 kg$\,$d.}
\label{inner-counts}
\begin{tabular}{lclc}
Bin [keV]&Counts/keV&Bin [keV]&Counts/keV\\\hline
2-3 & 1& 25-26 & 1 \\
3-4 & 2 & 26-27 & 1 \\
4-5 & 7 & 28-29 & 3 \\
5-6 & 6 & 29-30 & 0\\
6-7 & 11 & 30-31 &0 \\
7-8 & 8 &31-32 & 0 \\
8-9 & 12 &32-33 &1 \\
9-10 & 38  &33-34 &3 \\
10-11 & 37 &34-35 &4 \\
11-12 & 1 &35-36 &4 \\
12-13 & 1 &36-37 &1 \\
13-14 & 1 &37-38 &1 \\
14-15 & 2 &38-39 &3 \\
15-16 &1 &39-40 &2 \\
16-17 & 0 &40-41 &1 \\
17-18 &1 &41-42 &0 \\
18-19 & 0 &42-43 &2 \\
19-20 & 3 &43-44 &1 \\
20-21 &1 &44-45 & 0 \\
21-22 &0 &45-46 &0 \\
22-23 &0 &46-47 &1 \\
23-24 &0 &47-48 &0 \\
24-25 &1 &48-49 &2 \\
\end{tabular}
\end{table}

\begin{table}
\caption{List of parameters used for calculating WIMP spectra.}
\begin{tabular}{ll}
Parameter&Value\\\hline
WIMP velocity distribution& 270 km/s\\
Escape velocity& 580 km/s\\
Earth velocity& 245 km/s\\
WIMP local density& 0.3 GeV/cm$^{3}$\\
natural Ge mass& 72.61 g/mol\\
\end{tabular}
\label{tab:parameter}
\end{table}

\newpage 

\end{document}